# Neutron Therapy in the 21st Century

**Thomas K Kroc, PhD[1], James S Welsh, MS, MD[2]**
[1]Fermi National Accelerator Laboratory, Batavia, Il, USA
[2]Northern Illinois University, DeKalb, Il, USA

**Abstract**

The question of whether or not neutron therapy works has been answered. It is a qualified yes, as is the case with all of radiation therapy. But, neutron therapy has not kept pace with the rest of radiation therapy in terms of beam delivery techniques.  Modern photon and proton based external beam radiotherapy routinely implements image-guidance, beam intensity-modulation and 3-dimensional treatment planning. The current iteration of fast neutron radiotherapy does not. Addressing these deficiencies, however, is not a matter of technology or understanding, but resources.

   The future of neutron therapy lies in better understanding the interaction processes of radiation with living tissue. A combination of radiobiology and computer simulations is required in order to optimize the use of neutron therapy. The questions that need to be answered are: Can we connect the macroscopic with the microscopic? What is the optimum energy? What is the optimum energy spectrum? Can we map the sensitivity of the various tissues of the human body and use that knowledge to our advantage? And once we gain a better understanding of the above radiobiological issues will we be able to capitalize on this understanding by precisely and accurately delivering fast neutrons in a manner comparable to what is now possible with photons and protons?

   This presentation will review the accomplishments to date. It will then lay out the questions that need to be answered for neutron therapy to truly be a 21st Century therapy.

**Introduction**

Fermilab's first director, Robert Wilson, first proposed hadron therapy in 1946 [1]. Fast neutron therapy began in the United States through a series of grants in the early 1970s from the National Cancer Institute to approximately eight facilities around the country [2]. In most cases, each facility was an add-on to an existing physics laboratory and used the beam and beam energy peculiar to that facility. Interest in offering neutron therapy at Fermilab grew out of a presentation by Louis Rosen, of the Los Alamos Scientific Laboratory, at the 1971 Particle Accelerator Conference and subsequent discussions among Chicago area physicians and physicists [3].

After Rosen's talk, local physicians and physicists began discussions to start hadron therapy in the Chicago area. Prof. Lester Skaggs of the University of Chicago and the Argonne Cancer Hospital organized these discussions looking into protons, ions, and pions. It is unclear what caused the shift from these particles to neutrons, but on September 7, 1976, the first patient was treated at the Cancer Therapy Facility, later named the Neutron Therapy Facility (NTF), at Fermilab with neutrons.

**Why Neutrons**

The primary reason supporting the use of neutrons for therapy is their relative biological effectiveness (RBE). For the neutron energies supplied by the NTF beam, 1/3 less dose is required to achieve the same clinical effect with neutrons as is required with conventional photons. However, this is not the complete story as there is probably no clinical advantage to using such a beam if the final outcome were exactly the same as with photons. Certain tumours are classified as being radioresistant. They respond very poorly to conventional photon therapy. In these cases, neutrons are more effective, beyond just the factor of three in RBE (see Table 1)[4]. A partial explanation is that conventional radiation therapy relies on the creation of oxygen free radicals to provide the lethal effect. Radioresistant tumours tend to be hypoxic which inhibits the creation of radicals. Neutrons on the other hand do not rely on radicals and therefore are less dependent on the oxygenation of the tumour. But even this does not completely explain a neutron's RBE. Other mechanisms are at work that are not completely understood.

In addition to fast neutron therapy, a new opportunity is appearing, neutron capture therapy (NCT). This is a binary therapy where a neutron absorbing agent is attached to a drug that is preferentially taken up by tumour cells. When exposed to neutron radiation, the agent absorbs a neutron and undergoes a radioactive decay. The energetic by-products of the decay provide a localized boost of dose to the tumour area in addition to the dose from the neutron radiation itself. The two agents presently being investigated are boron-10 (BNCT) and two isotopes of gadolinium, 155 and 157, (GdNCT).

**Table 1: Review of the loco-regional rates for malignant salivary gland tumors treated with radiation therapy.**

| Fast Neutrons | | | |
|---|---|---|---|
| Authors | Number of Patients | Loco-regional control (%) | |
| Saroja *et al.* (1987) | 113 | 71 | (63%) |
| Catterall and Errington (1987) | 65 | 50 | (77%) |
| Battermann and Mijnheer (1986) | 32 | 21 | (66%) |
| Griffin *et al.* (1988) | 32 | 26 | (81%) |
| Duncan *et al.* (1987) | 22 | 12 | (55%) |
| Tsunemoto *et al.* (1989) | 21 | 13 | (62%) |
| Maor *et al.* (1981) | 9 | 6 | (67%) |
| Ornitz *et al.* (1979) | 8 | 3 | (38%) |
| Eichhorn (1981) | 5 | 3 | (60%) |
| Skolyszewski (1982) | 3 | 2 | (67%) |
| Overall | 310 | 207 | (67%) |

| Low-LET Radiotherapy Photon and/or Electron beams and/or Radioactive Implants | | | |
|---|---|---|---|
| Authors | Number of Patients | Loco-regional control (%) | |
| Fitzpatrick and Theriault (1986) | 50 | 6 | (12%) |
| Vikramet *et al.* (1984) | 49 | 2 | (4%) |
| Borthne et al. (1986) | 35 | 8 | (23%) |
| Rafla (1977) | 25 | 9 | (36%) |
| Fu *et al.* (1977) | 19 | 6 | (32%) |
| Stewart *et al.* (1968) | 19 | 9 | (47%) |
| Dobrowsky *et al.* (1986) | 17 | 7 | (41%) |
| Shidnia *et al.* (1980) | 16 | 6 | (38%) |
| Elkon *et al.* (1978) | 13 | 2 | (15%) |
| Rossman (1975) | 11 | 6 | (54%) |
| Overall | 254 | 61 | (24%) |

Table III. from IAEA-TECDOC-992, "Nuclear data for neutron therapy: Status and future needs," December 1997, pg. 12.

**The Fermilab Setup**

At Fermilab, the neutron beam is produced by bombarding a beryllium target with protons. The Neutron Therapy Facility is located approximately a third of the length of the proton linac used as the injector for Fermilab's High Energy Physics program. As the linac pulses at 15 Hz, a fast 58° dipole magnet can divert linac pulses into the NTF beam line. A second 32° dipole magnet completes the 90° bend. Quadrupole magnets maintain the focus of the beam in the 12 foot long beam line and direct the beam onto the target. At the point where the beam is diverted from the linac, the proton energy is 66 MeV. The beryllium target is 49 MeV thick or 2.21 cm and is 2.5 cm in diameter (see Figure 1). It is backed by 0.05 cm of gold and placed in an aluminium holder which is 0.32 cm thick in the beam direction. The latter two elements ensure that all the protons have ranged out so that only photons and neutrons escape the target.

A primary collimator of steel with a conical opening of 7.5° is immediately downstream of the target assembly. This followed by a transmission chamber which is used to monitor the flux of neutrons generated. The neutrons then enter the therapeutic collimator assembly. This contains interchangeable collimators of either a concrete/polyethylene matrix or polyurethane that have various rectangular openings. The proper choice of one of these collimators determines the size of the field to be used for treatment. These collimators are 78 cm long. To better conform the outline of the neutron field to the shape of the tumour to be treated, blocks, 20 cm long, made from low carbon steel are placed in the opening of the collimator.

**Figure 1: NTF Target and Collimation**

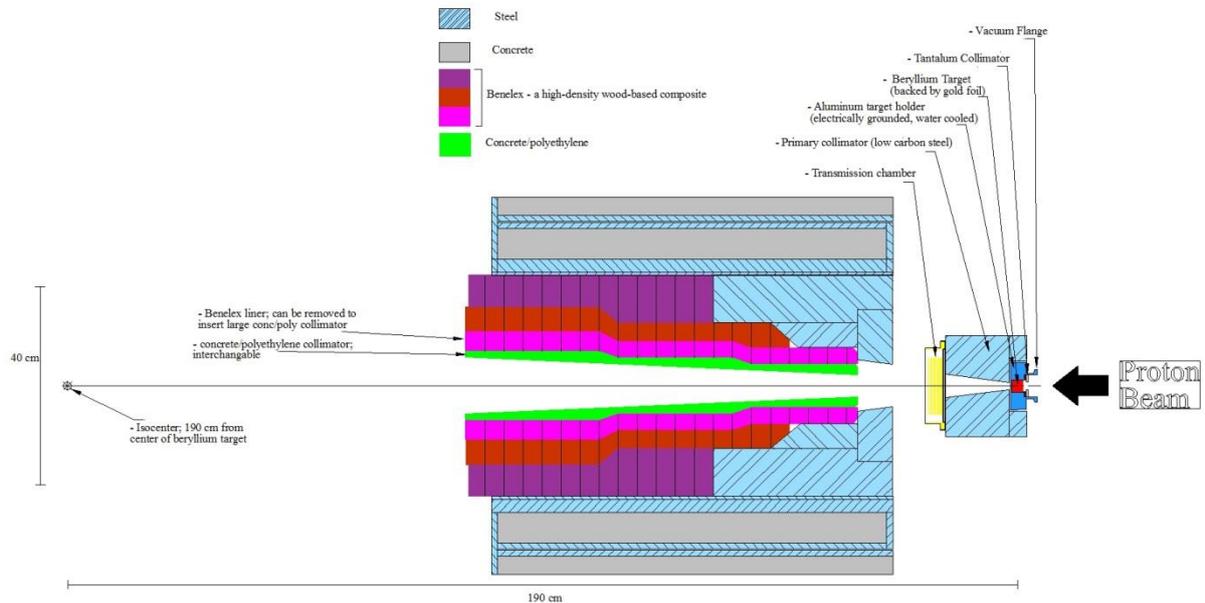

While the beam is fixed in the horizontal plane, treatment is able to be delivered in an isocentric manner by moving the patient. The patient either stands or sits on an immobilization platform that rotates in front of the beam opening. The platform can translate along, and transversely to, the beam axis allowing the placement of the tumour above the centre of rotation of the platform. This allows the tumour to be irradiated from multiple angles throughout the treatment while remaining fixed at the isocentric distance of 190 cm.

**A Matter of Scale**

While trying to understand how neutrons, and all radiation for that matter, truly interact to cause their damage, one is faced with the drastic range of distance scale involved. This is apparent in both conducting physical measurements and in simulations. Macroscopic measurements, with distance scales of centimetres or millimetres are easily achievable in both cases.

The conventional wisdom is that linear energy transfer (LET), in the context of double-strand breaks in the DNA, is the key to understanding the killing power of neutrons. This wisdom, in effect, assumes that LET is a linear value that can somehow be varied over orders of magnitude with an optimum value at 100 keV/µm. This value of LET happens to match well with the distance scale of the double helix of DNA. Despite a criticism that will be mentioned shortly, this means that we need to understand the interactions of fast neutrons and matter on a distance scale of a few nanometres. This is in contrast with the distance scale of BNCT where the range of fission products of the boron atom is 7.3 µm and 4 µm for the alpha and lithium ions respectively, whereas the range of the Auger electrons from gadolinium decay is again a few nanometres.

**The Connection to Chemistry and Biology**

It is becoming apparent that simple DNA breakage is not sufficient in understanding the killing power of ionizing radiation. A broader understanding of the ways ionizing radiation can disrupt cellular processes is needed [5]. We previously mentioned RBE as a factor in determining the need for neutron therapy along with LET and double-strand breaks. However, this understanding, particularly of LET, is not completely adequate for describing the impact of neutrons. Radiobiologist Shirley Lehnert provides this critique of LET: "LET is a simplistic way to describe the quality of different types of radiation since it fails to address the size of the individual energy-loss events that occur along the track of a particle." [6] Understanding "the size of the individual energy-loss events" could very well be the key in understanding the difference between photon, neutron, and light-ion initiated events. Simulations of the radiation/matter interactions on scales comparable to the biological processes involved will be necessary to truly understand and exploit ionizing radiation.

**The Role of Simulations**

In this paper we only report on our own efforts in simulating the NTF beam. MCNPX [7] was chosen because of past experience with the style of input format used in operating the code. Very early on we had the benefit of comparing some initial results with GEANT 4 [8][9] and that work served to validate our initial efforts and give us some assurance that we were operating the code correctly (Figure 2).

**Figure 2: Comparison of MCNPX and GEANT4 results showing proton energies after passing through beryllium and gold of target assembly**

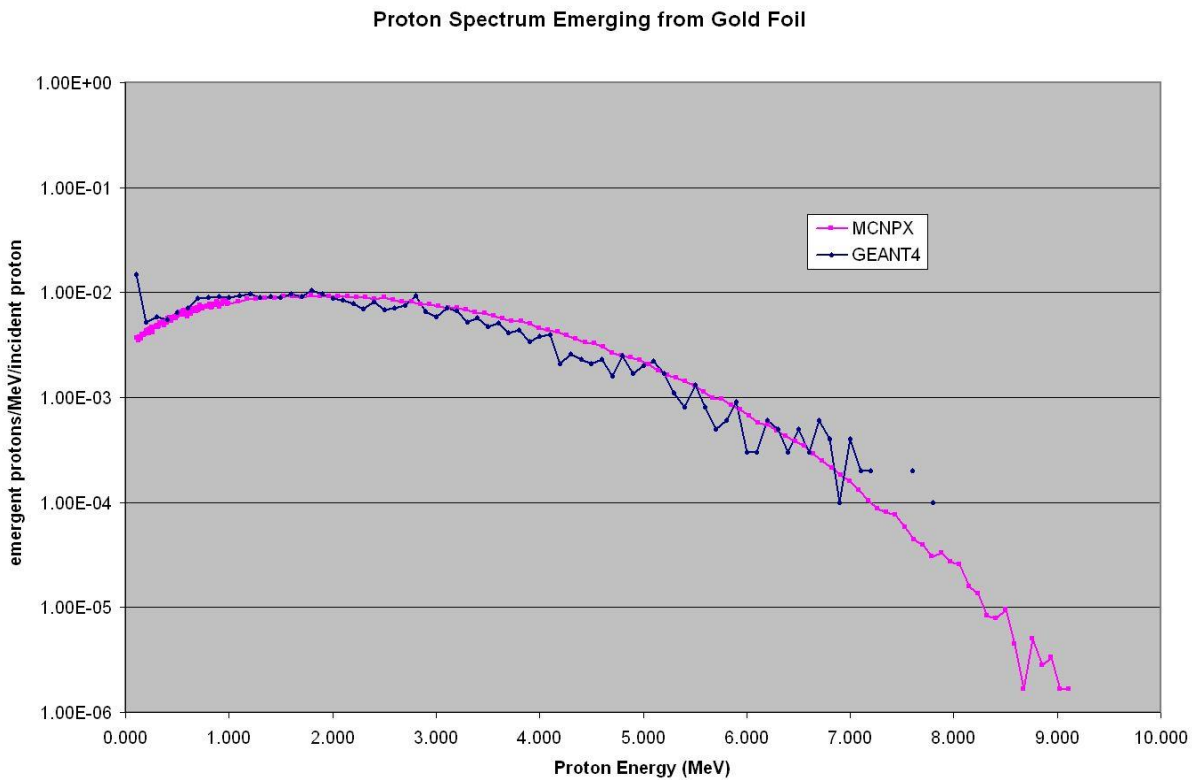

Our work with MCNPX has been useful in understanding the macroscopic evolution of the neutron energy as it passes through the system. Much of our work has been focused on finding ways to moderate our fast neutron beam so that it is better suited to NCT. We have tried various filtration schemes to try to

maximize the production of epithermal neutrons (Figure 3). This figure shows
the neutron energy spectrum at the exit of the clinical collimation system
(yellow). After exiting the collimator the beam is then simulated to pass
through either 10 cm of lead (purple) or 10 cm of tungsten (dark blue).
Alternately, the figure shows the results of the neutron passing through
alternating 1 cm thicknesses of lead and tungsten, with a total of 5 cm each
(cyan). The goal of this work is to generate a neutron spectrum that retains
enough penetrating power to reach deep tumours and yet moderate in energy as
they pass through tissue so that the energy is low enough for capture once they
encounter the enhancing compound in the tumour. However, we are not able to see
what happens at scales smaller than a few millimetres.

**Figure 3: Simulation of neutron spectra leaving collimator system (input spectra) and after passing through 4 cm of lead and/or tungsten to optimize neutron production in epithermal range.**

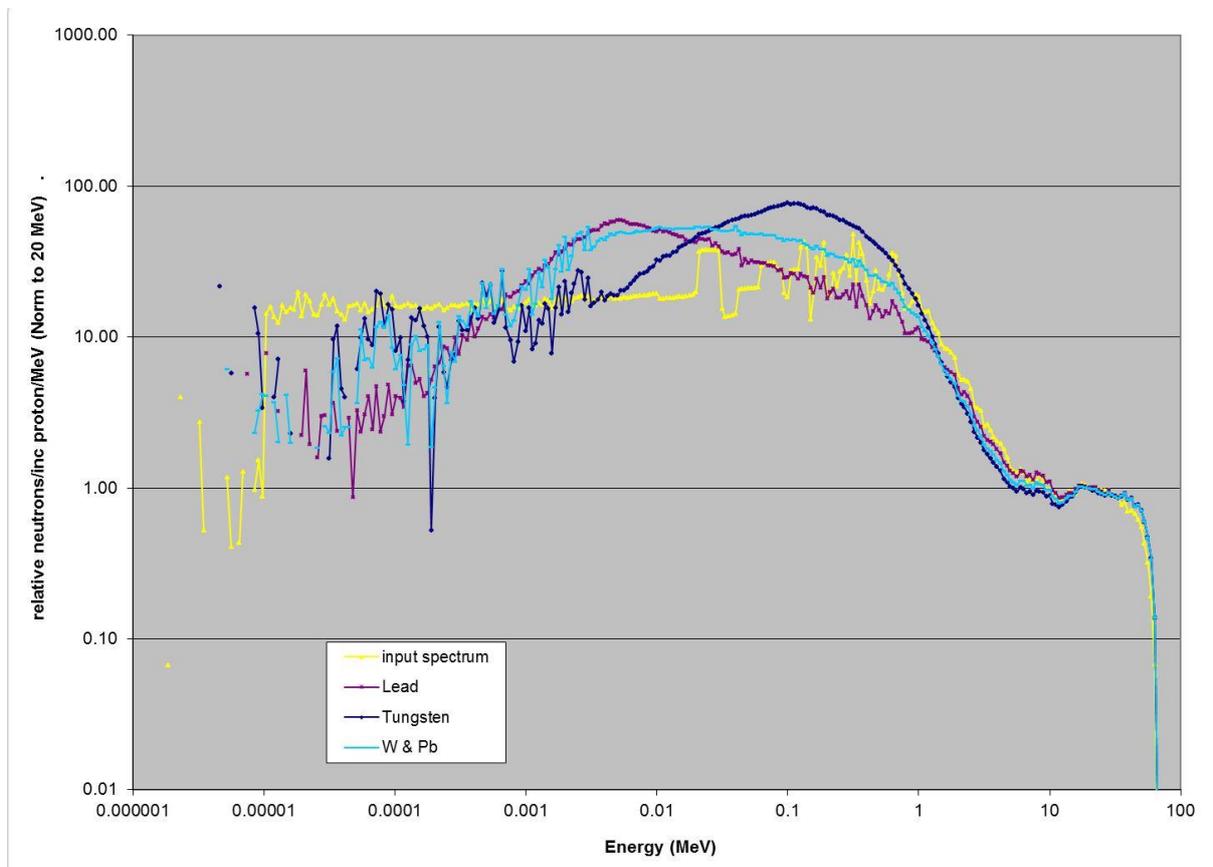

As mentioned above, a challenge in NCT is having enough penetrating power
to reach deep tumours. Figure 4 shows the range of neutrons of various initial
energies (Figure 4). This shows that neutrons with initial energies below 100
keV cannot contribute to a clinical dose at depth. This has implications for
both fast neutron therapy and NCT. For fast neutron therapy, it would be
advantageous to remove these neutrons from the incident beam as they only
contribute to higher risk for complications. For NCT, it places a limit on the
lower edge of the energy spectrum that we hope will moderate, as it penetrates
tissue, to near thermal energies by the time they reach the tumour.

**Figure 4: The 1/e attenuation distance of mono-energetic neutrons striking A150 plastic.**

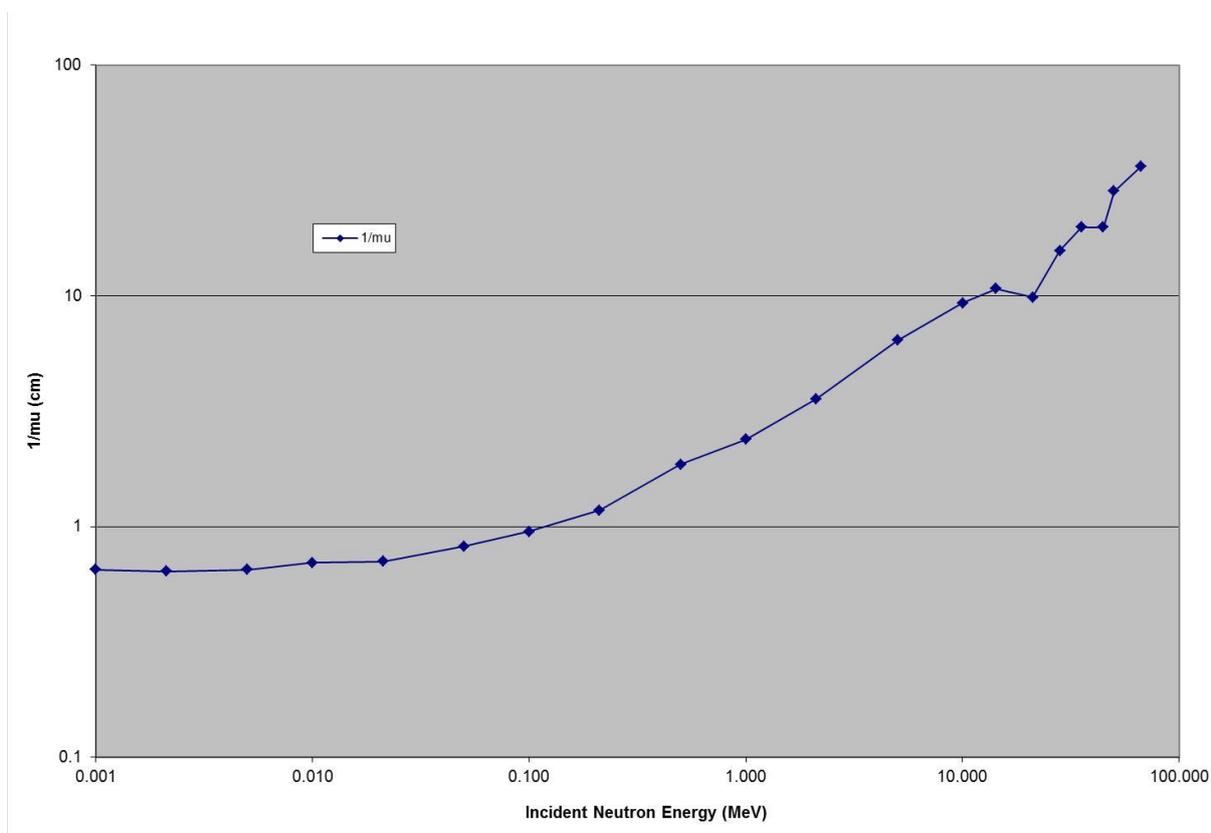

**Future Needs**

There are many applications waiting for accurate simulations of the interaction of radiation with matter at the nanometer level. Physics, chemistry and biology can all benefit from this development. The effective treatment of cancer and other diseases need an understanding of the magnitude and type of interactions that occur. Are certain chromosomal structures more or less susceptible to damage? What kind of radiation induces this damage? How do the interactions of the recoil fission products of BNCT interact differently than the dense Auger electron shower of GdNCT? Do fast neutrons produce different secondary ionizing radiation depending on their energy?

Once answered, this knowledge can be entered into optimization routines for treatment planning systems. Instead of plotting dose deposition, actual killing power can be visualized. Target structures can be enhanced and sensitive tissues can be avoided.

Then we will truly have a 21$^{st}$ Century therapy.

**Acknowledgements**